\documentclass[fleqn,twocolumn,10pt]{revtex4}

\usepackage{amsmath}
\usepackage{amssymb}
\usepackage{graphicx,psfrag}
\usepackage{afterpage}
\usepackage{mathrsfs}

\newcommand{\bwt}{\begin{widetext}}
\newcommand{\ewt}{\end{widetext}}
\newcommand{\be}{\begin{equation}} \newcommand{\ee}{\end{equation}}
\newcommand{\bea}{\begin{eqnarray}} \newcommand{\eea}{\end{eqnarray}}
\newcommand{\bean}{\begin{eqnarray*}} \newcommand{\eean}{\end{eqnarray*}}
\newcommand{\bm}[1]{\mbox{\boldmath $#1$}}
\newcommand{\s}[1]{{\scriptscriptstyle #1}}
\newcommand{\sT}{{\s T}}

\newcommand{\nn}{\nonumber \\}

\begin{document}

\title{Spectral analysis of gluonic pole matrix elements for fragmentation}

\author{L. P. Gamberg}
\email{lpg10@psu.edu}
\affiliation{Division of Science, 
Penn State University  Berks Campus, Reading,Pennsylvania 19610, USA}

\author{A. Mukherjee}
\email{asmita@phy.iitb.ac.in}
\affiliation{Physics Department,
Indian Institute of Technology, Powai, Mumbai 400076,
India}

\author{P.J. Mulders}
\email{mulders@few.vu.nl}
\affiliation{
Department of Physics and Astronomy, VU University\\
NL-1081 HV Amsterdam, the Netherlands}

\begin{abstract}
The non-vanishing of gluonic pole matrix elements can explain the
appearance of single spin asymmetries in high-energy scattering processes.
We use a spectator framework approach to investigate the spectral properties
of quark-quark-gluon correlators and use this to study gluonic pole 
matrix elements. Such matrix elements appear in principle both for 
distribution functions such as the Sivers function and fragmentation 
functions such as the Collins function. We find that for a large class of 
spectator models, the contribution of the gluonic pole matrix element in
fragmentation functions vanishes. This outcome is important in the 
study of universality for fragmentation functions and confirms findings
using a different approach.
\end{abstract}

\date{\today}


\maketitle

\section{Introduction}
In high-energy scattering processes the structure of hadrons is 
accounted for using quark and gluon correlators, forward matrix elements
of non-local quark and gluon operators between hadronic states. Making an
expansion in the (inverse) hard scale, the relevant component of the momentum
of partons (quarks and gluons) is the one collinear to hadrons 
and correspondingly 
the non-locality in the matrix elements is restricted to the
light-cone. Moreover, all leading dynamical effects come from two-field
configurations at two light-like separated points,
which are easily interpreted as parton densities or parton decay
functions~\cite{Collins:1981uw,Jaffe:1991ra}. These are the parton
distribution functions depending on the momentum fraction $x$ relating 
the parton momentum $k = x\,P$ to the hadron momentum $P$
or the fragmentation functions of partons into hadrons depending
on the momentum fraction $z$, relating the parton momentum $k$ and
the hadron momentum $P = z\,k$. 

At sub-leading order in the hard scale or when explicitly measuring 
transverse momenta, other matrix elements become important such as 
the three-parton correlators containing parton fields
at three different space-time points with light-like separations
and two-parton correlators with also transverse separation (light-front
correlations). These latter (light-front) correlators are described
in terms of transverse momentum dependent (TMD) distribution and 
fragmentation functions, 
which are  sensitive to the intrinsic transverse momenta
of partons in hadrons, $k = x\,P + k_\sT$ in a frame in which the
hadron does not have transverse momentum ($P_\sT = 0$) or for
fragmentation $k = \tfrac{1}{z}\,P + k_\sT$. In this case one
often refers to the hadron transverse momentum $P_\perp = -z\,k_\sT$ 
(in a frame in which the parton does not have a transverse momentum
($k_\perp = 0$)).
\begin{figure}
\begin{center}
\includegraphics[width=7cm]{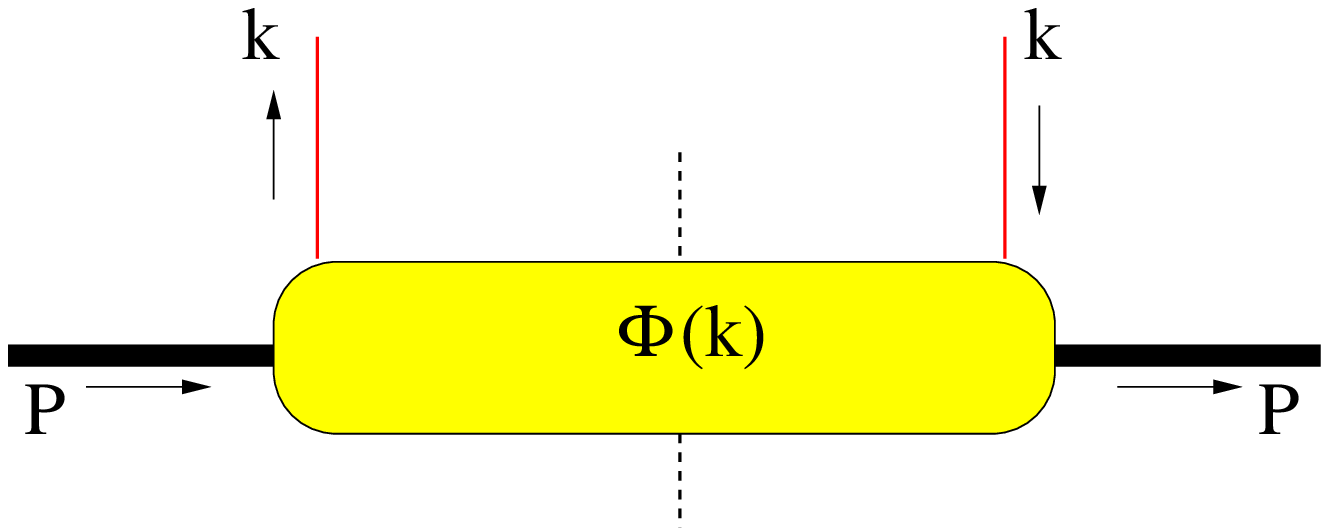}\\ 
\vspace{.2cm}
(a)\\
\vspace{.2cm}
\includegraphics[width=7cm]{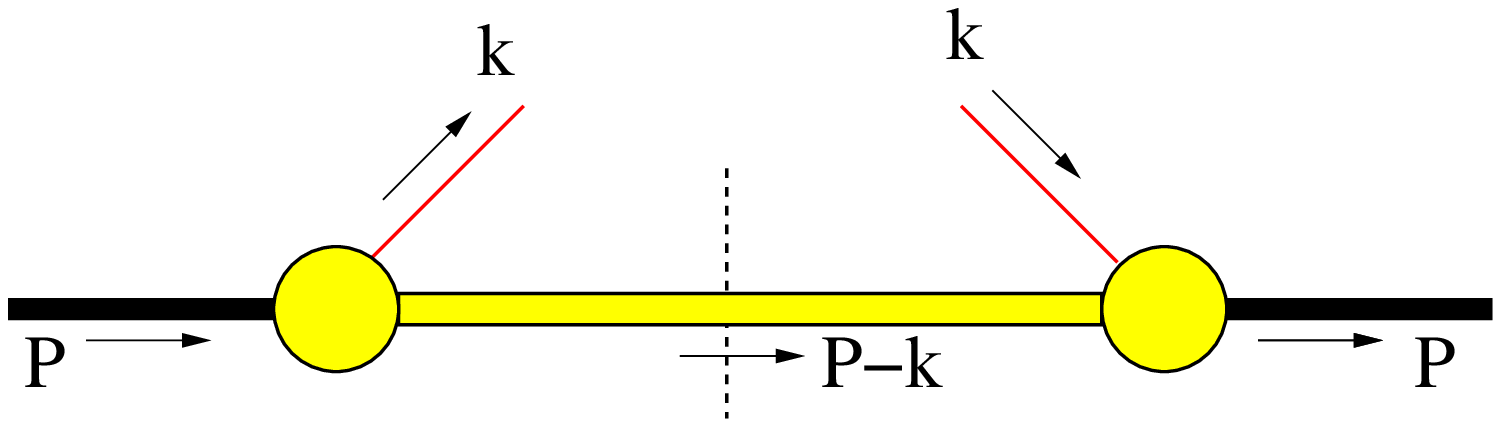}\\
\vspace{.2cm}
(b)
\end{center}
\caption{\label{qqspec-1}
The graphical representation of the quark-quark correlator in the case of
distributions of partons with momentum $k$ in a hadron with
momentum $P$ (a) and the spectator model description (b).}
\end{figure}

\begin{figure}
\begin{center}
\includegraphics[width=3.9cm]{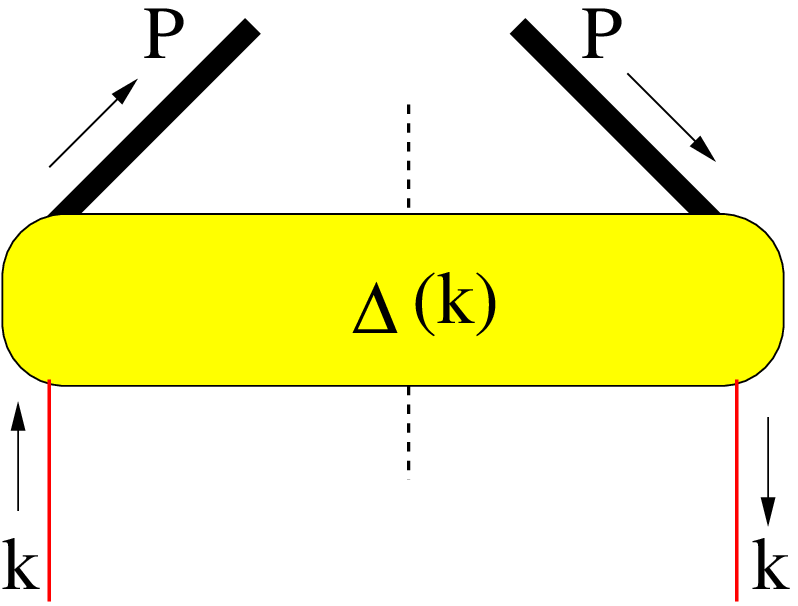}\\
\vspace{.2cm} (a)\\ \vspace{.2cm}
\includegraphics[width=6cm]{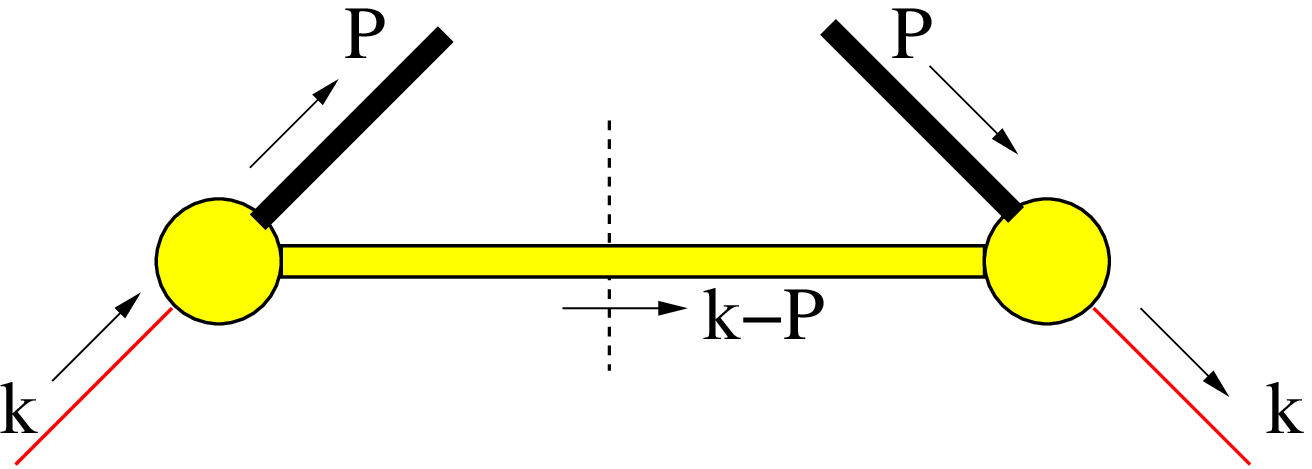}\\
\vspace{.2cm} (b)
\end{center}
\caption{\label{qqspec-2}
The graphical representation of the correlator in the case of
fragmentation of partons with momentum $k$ into a hadron with
momentum $P$ (a) and the spectator model description (b).}
\end{figure}

In this paper, we will investigate multi-parton
correlators with one additional gluon in which the zero-momentum
limit will be studied~\cite{Efremov:1981sh,Efremov:1984ip,Qiu:1991pp,Qiu:1991wg,Qiu:1998ia,Kanazawa:2000hz,Eguchi:2006mc,Koike:2006qv}. 
These are so-called gluonic pole matrix
elements or Qiu-Sterman matrix elements, that have opposite 
time-reversal (T) behavior as compared to
the matrix elements without the gluon. Such matrix elements involving 
time-reversal odd (T-odd) operator combinations are of interest because
they are essential for understanding single spin asymmetries 
at high energies. 
 In order
to understand the basic features of these matrix elements we 
perform a spectral analysis by modeling the distribution and
fragmentation functions under {\em reasonable} assumptions.  

\begin{figure}
\begin{center}
\includegraphics[width=7cm]{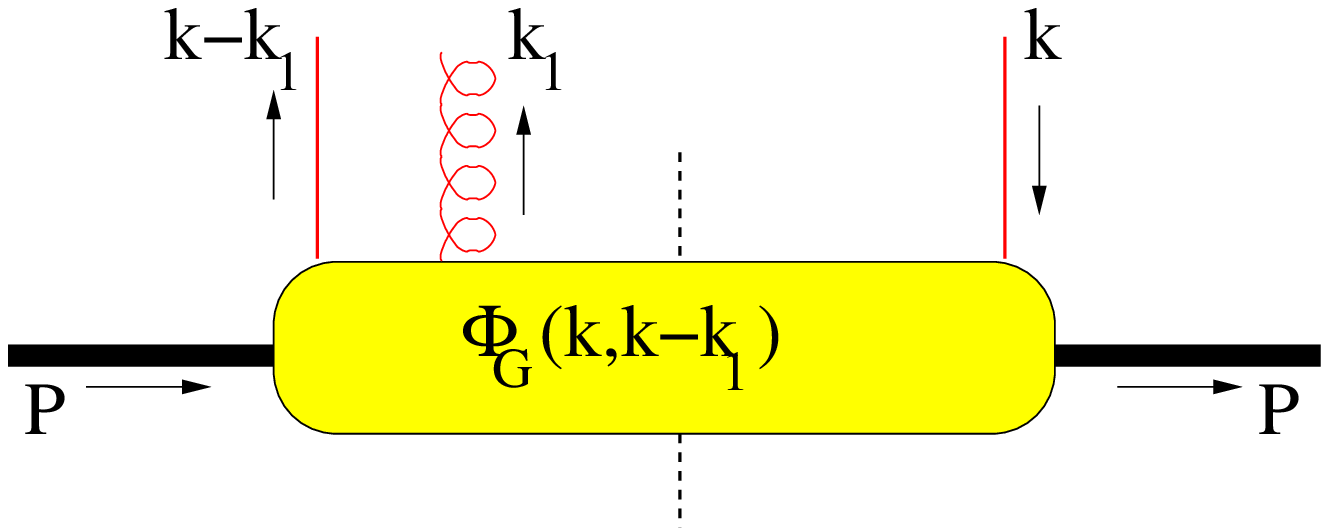}\\
\vspace{.2cm}(a)\\ \vspace{.2cm}
\includegraphics[width=7cm]{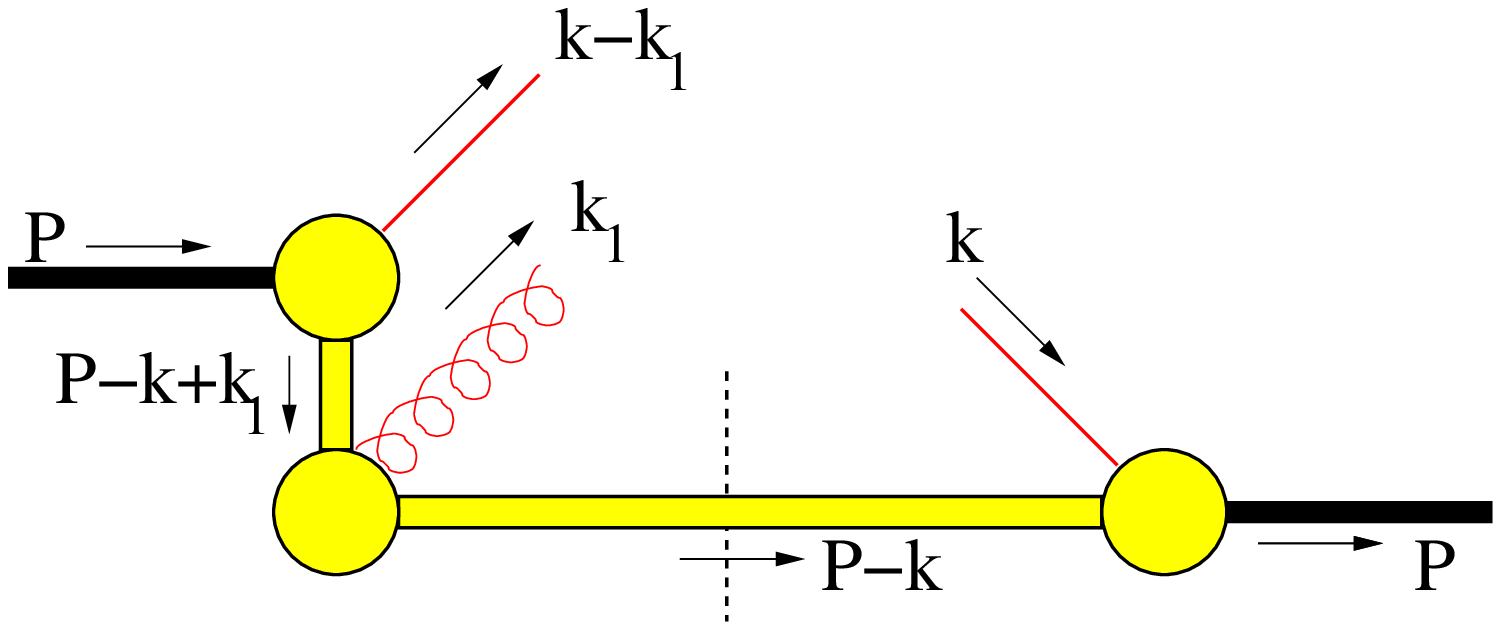}\\
\vspace{.2cm}(b)\\ \vspace{.2cm}
\includegraphics[width=7cm]{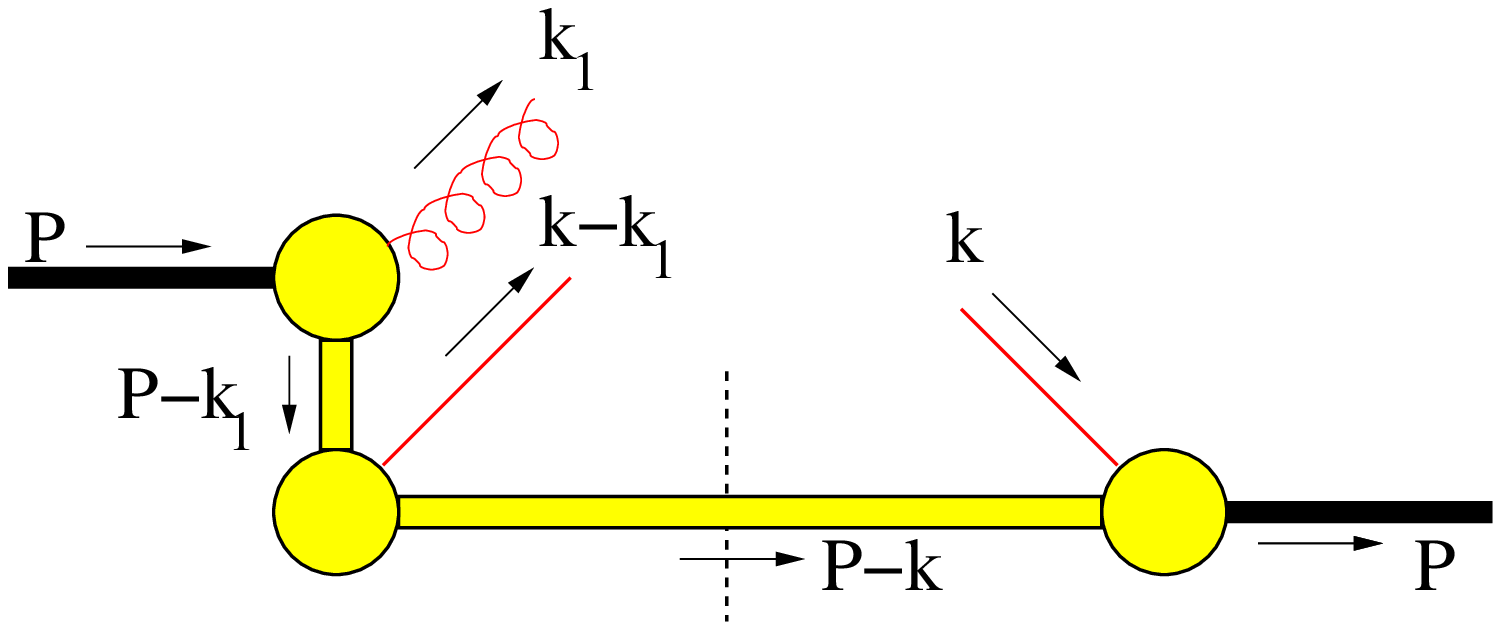}\\
\vspace{.2cm}(c) \\ \vspace{.2cm}
\end{center}
\caption{\label{qqGspec-1}
The graphical representation of the quark-quark-gluon
correlator $\Phi_G$ for  the case of distributions including
a gluon with momentum $k_1$ (a), and the
possible intermediate states (b) and (c) in a spectator 
model description.}
\end{figure}

\begin{figure}
\begin{center}
\includegraphics[width=4.5cm]{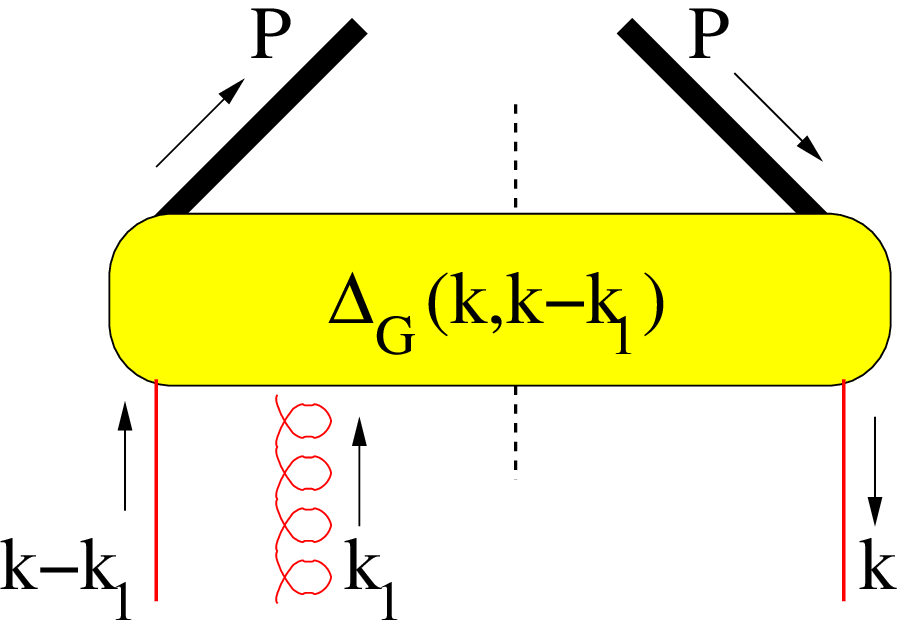}\\
\vspace{.2cm}(a)\\ \vspace{.2cm}
\includegraphics[width=6cm]{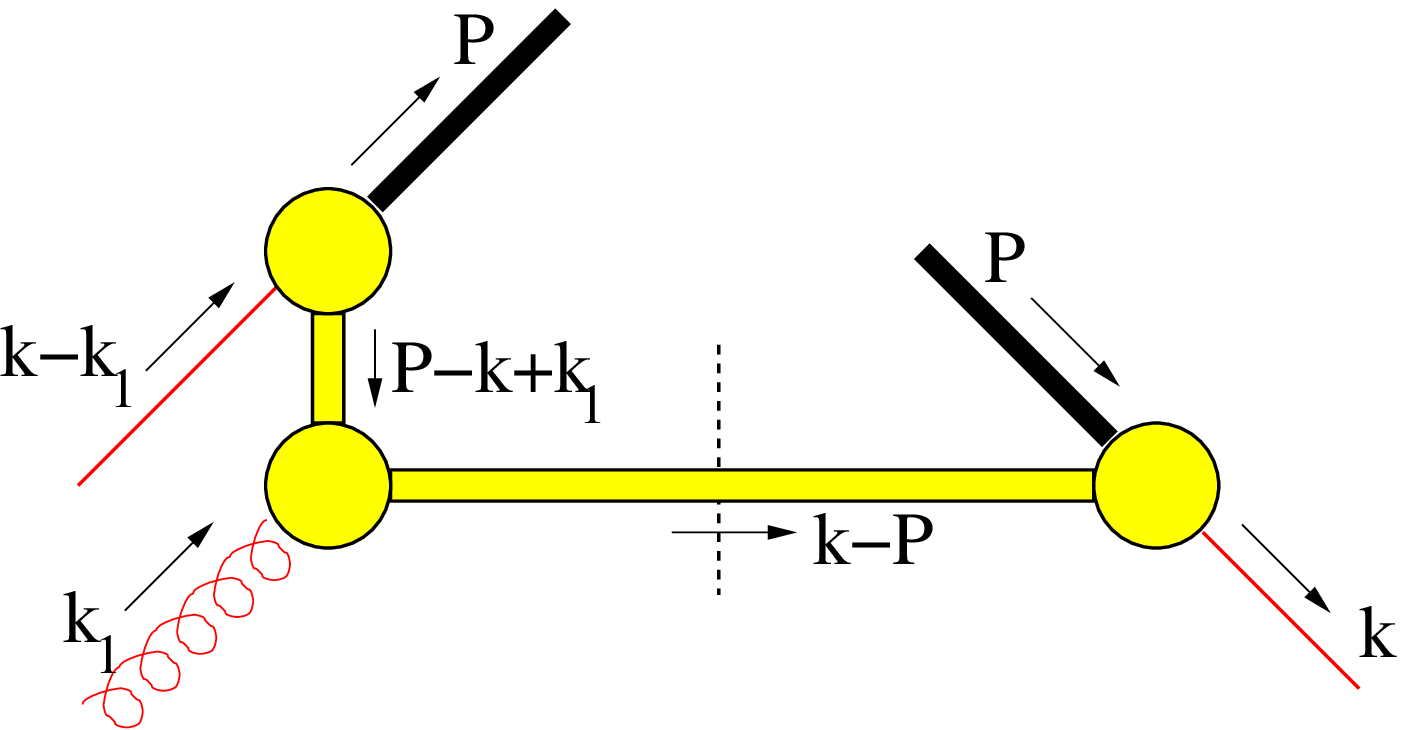}\\
\vspace{.2cm}(b)\\ \vspace{.2cm}
\includegraphics[width=6cm]{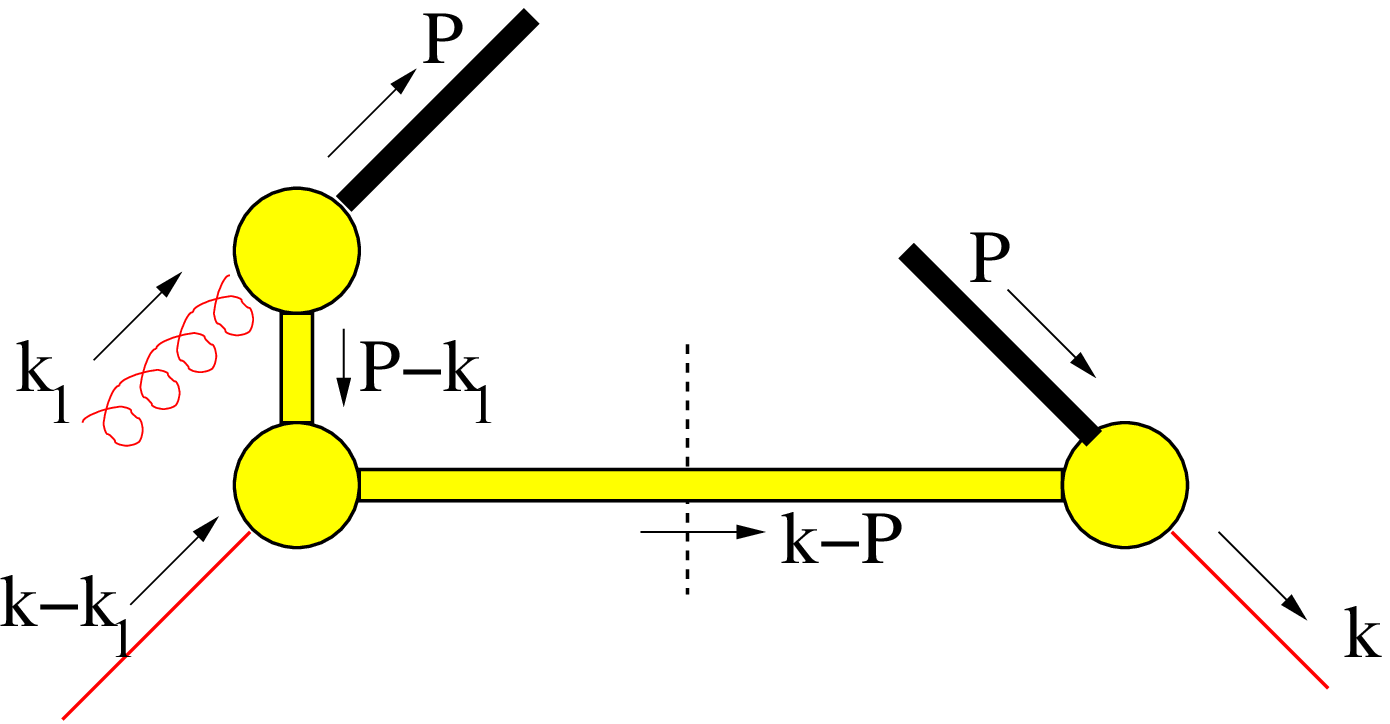}\\
\vspace{.2cm}(c)\\ \vspace{.2cm}
\end{center}
\caption{\label{qqGspec-2}
The graphical representation of the quark-quark-gluon
correlator $\Delta_G$ in the case of fragmentation including
a gluon with momentum $k_1$ (a) and the
possible intermediate states (b) and (c) in a spectator 
model description.}
\end{figure}

For the correlators, depicted in Figs.~\ref{qqspec-1}(a)
and \ref{qqspec-2}(a), one has expressions in terms of matrix elements 
of bilocal operators that are frequently used as a  starting point in 
modeling distribution and fragmentation functions. 
In particular the spectator model, pictorially given
in Figs.~\ref{qqspec-1}(b) and \ref{qqspec-2}(b), has become
fairly popular, because it is easy, flexible and intuitively
attractive. On the other hand, one should be very careful, because
the predictive power depends on limiting oneself in the choice of
spectator (e.g.\ a diquark with fixed mass in case of the nucleon) and 
using simple vertices. In fact making a spectral analysis of
the spectator and allowing for the most general vertices one
would lose all predictive power. Having said these words of caution,
we will investigate in this paper differences between distribution
and fragmentation functions using a spectral analysis and using
physical intuition in restricting the momentum dependence and
asymptotic behavior of the vertices.   
In this context, the relevant gluonic pole matrix elements 
that we want to  study are 
$\Phi_G(k,k-k_1)$ and $\Delta_G(k,k-k_1)$ shown in 
Figs.~\ref{qqGspec-1} and \ref{qqGspec-2}. Of these matrix
elements only the dependence on the collinear components
$x$ and $x_1$ in the expansion of the momenta is needed.
We find that while both $\Phi_G(x,x-x_1)$
and $\Delta_G(x,x-x_1)$ are nonzero,
taking the limit $x_1 \rightarrow x$, $\Phi_G(x,x)$ 
remains non-zero, while $\Delta_G(x,x)$ vanishes.

The vanishing of the T-odd gluonic pole matrix elements is important
in the study of universality of TMD distribution functions (DFs) and 
fragmentation functions (FFs). 
For the collinear case T-symmetry can be used as a constraint
on the parton correlators, limiting the DFs to T-even ones. Such a 
constraint does not apply for the fragmentation correlator because
the final state hadron is part of a jet and as such not a plane wave, 
allowing both T-even and T-odd FFs. 
But for spin 0 and spin 1/2 hadrons no T-odd 
functions appear at leading 
twist (leading order in the hard scale)~\cite{Ji:1993vw,Bacchetta:2000jk}.
Including transverse momentum dependence, both the distribution and
fragmentation correlators ($\Phi$ and $\Delta$) are
no longer constrained by T-symmetry. The reason is that the
appropriate color gauge invariant operators in the correlator,
in particular the gauge links, are
not T-invariant. For DFs this provides a mechanism leading to T-odd
functions, such as the Sivers function~\cite{Sivers:1989cc}. But, for FFs there are now
in principle two mechanisms leading to T-odd functions~\cite{Boer:2003cm}. 

A nice feature, however, is that the two mechanisms leading to 
T-odd functions can be distinguished. The T-odd operator structure
can be traced back to the color gauge link that necessarily appears in
correlators to render them color gauge-invariant. But the operator
structure of the correlator is also a consequence of the necessary
resummation of all contributions that arise from collinear gluon
polarizations, i.e.\ those along the hadron momentum. How this 
resummation takes effect is a matter of calculation. The result is a 
process dependence in the path in the gauge link. 
After azimuthal weighting of cross sections one 
simply finds that the T-odd features originating from the gauge link
lead to specific factors with which the T-odd functions appear in 
observables. Comparing T-odd effects in DFs in
semi-inclusive deep inelastic scattering (SIDIS) and the Drell-Yan 
process one finds a relative minus sign~\cite{Collins:2002kn}. 
Similarly, comparing T-odd 
effects in FFs in SIDIS and electron-positron
annihilation one also finds a relative minus sign, at least for the
T-odd effect originating from the operator 
structure (gauge link)~\cite{Boer:2003cm}.
The effect coming from the hadron-jet final state not being a plane
wave will not lead to process dependent factors.

Before coming to the significance of our work, we need one more
ingredient. We already mentioned that relating T-odd effects in
different processes, requires azimuthal weighting, which projects
out the transverse momentum weighted parts of the correlators
$\Phi$ and $\Delta$, referred to as transverse moments $\Phi_\partial$ 
and $\Delta_\partial$, respectively. The T-odd operator parts
are precisely the soft limits ($k_1\rightarrow 0$) of the
gluonic pole matrix elements~\cite{Boer:2003cm}. Thus
\bea
&&\Phi_\partial = \tilde\Phi_\partial + \pi\Phi_G(k_1=0),
\\
&&\Delta_\partial = \tilde\Delta_\partial + \pi\Delta_G(k_1=0),
\eea
When $\Delta_G(k_1=0)$ is zero, there still are T-odd FFs 
contained in $\Delta_\partial$. They appear in the matrix elements
of the T-even operator combination in $\tilde \Delta_\partial$ involving
a hadron-jet state (non-plane-wave) and they are process independent, 
for instance the T-odd Collins function~\cite{Collins:1992kk}. 
In contrast T-odd DFs in
$\Phi_\partial$ only can come from $\Phi_G(k_1=0)$. These DFs can still
be universal but appear with calculable process dependent gluonic pole 
factors~\cite{Bomhof:2004aw,Bomhof:2006ra}.

The above had already been shown in model calculations for SIDIS and $e^+e^-$ 
annihilation~\cite{Metz:2002iz,Collins:2004nx} and more recently
in hadron-hadron scattering~\cite{Yuan:2007nd,Yuan:2008yv}. 
In these calculations the authors  look at the full process in the model, 
carefully studying the cuts, rather
than concentrating on the soft part only.
By contrast, we look at the soft part only, that is, the multi-parton
correlators $\Phi_G$ and $\Delta_G$.  In principle such an approach of
only looking at the soft part is also possible starting with TMD dependent 
two-parton correlators $\Phi$ and 
$\Delta$~\cite{Bacchetta:2002tk,Ji:2002aa,Gamberg:2003ey,Gamberg:2003eg,
Bacchetta:2003xn,Bacchetta:2003rz,Lu:2004hu,Amrath:2005gv,Bacchetta:2007wc,Gamberg:2007gb}. 
To generate  T-odd contributions for TMD functions  
in such an approach one 
needs to go at least to one-loop~\cite{Brodsky:2002cx} 
performing a perturbative expansion on the 
gauge link~\cite{Ji:2002aa,Gamberg:2003ey}.
Extraction of the {\em complete} T-odd gluonic pole  contribution
is difficult.  Our approach, starting directly with the 
color gauge invariant multi-parton correlator
(that is having  re-summed the gauge
link~\cite{Boer:2003cm}) 
has the advantage we can work with  tree-level matrix elements
and just  perform a spectral analysis to extract the T-odd gluonic
pole contributions.

In the next section we give details on the gluonic pole matrix
elements, followed by the spectator model approach. In the concluding
section, we outline the shortcomings of our approach and also 
discuss possibilities to use the approach for more detailed 
estimates for the T-odd FFs.

\vspace{-.25cm}
\section{Gluonic pole matrix elements\label{GPsection}}

At high energies, it is useful to make a Sudakov decomposition
of the momenta of active partons, $k=x\,P+\sigma\,n+k_\sT$. The 
Sudakov vector $n$ is an arbitrary light-like four-vector $n^2=0$ 
that has non-zero overlap $P\cdot n$ with the hadron's momentum $P$. We
will simply choose $P\cdot n = 1$.
In a hard process, the Sudakov vector incorporates the presence 
of other momenta that are hard with respect to the hadron under 
consideration, e.g.\ $n \approx P^\prime/P\cdot P^\prime$. 
We can now also work with light-cone coordinates. Including mass effects 
one would have $n_- = n$ and $n_+ = P - \tfrac{1}{2}\,M^2\,n$ and with
$k^\pm \equiv k\cdot n_\mp$ we have
\bea
&& k^+ = k\cdot n = x
\\&& k^- = k\cdot P - \tfrac{1}{2}\,xM^2 = \sigma + \tfrac{1}{2}\,xM^2.
\eea
Vectors in the transverse plane can be obtained from the transverse
projector, 
$g_\sT^{\mu\nu}=g^{\mu\nu}-n_+^{\{\mu}n_-^{\nu\}}$.

As the effects of the component $k^-$ will appear suppressed by two
powers of the hard scale as compared to the collinear term, it is
integrated over and one considers quark-quark correlators 
on the light-front
(LF: $\xi\cdot n=0$)
\bea\label{TMDDF}
\Phi_{ij}^{[\mathcal U]}(x{,}k_\sT)
&=&{\int}\frac{d(\xi{\cdot}P)\,d^2\xi_\sT}{(2\pi)^3}\ e^{ik\cdot\xi}
\nn && \times\langle P|\,\overline\psi_j(0)\,\mathcal U_{[0;\xi]}\,
\psi_i(\xi)\,|P\rangle\big\rfloor_{\text{LF}}\ .
\eea
The \emph{Wilson line} or \emph{gauge link}
$\mathcal U_{[\eta;\xi]}
=\mathcal P{\exp}\big[{-}ig{\int_C}\,ds{\cdot}A^a(s)\,t^a\,\big]$
is a path-ordered exponential along the integration path $C$ with
endpoints at $\eta$ and $\xi$.
Its presence in the hadronic matrix element is required by gauge-invariance.
In the TMD correlator~\eqref{TMDDF} the integration path $C$ in
the gauge link is process-dependent.
In the diagrammatic approach the Wilson lines arise by resumming all
gluon interactions between the soft and the hard partonic parts of the
hadronic process~\cite{Efremov:1978xm,Boer:1999si,Belitsky:2002sm,Boer:2003cm}.

Collinear quark distribution functions are obtained from the TMD correlator
after integration over $p_\sT$,
\bea\label{colcorrelator}
\Phi(x)&=& \int d^2k_\sT\ \Phi^{[\mathcal U]}(x{,}k_\sT)
\nn
&=&{\int}\frac{d(\xi{\cdot}P)}{2\pi}\ e^{i\,x\,\xi\cdot P}\,
\langle P|\,\overline\psi(0)\,\mathcal U_{[0;\xi]}^n\,
\psi(\xi)\,|P\rangle\big\rfloor_{\text{LC}}\ .
\nn
\eea
The non-locality is restricted to the light-cone 
(LC: $\xi\cdot n = \xi_\sT=0$) and the gauge link is
unique, being the straight-line path along $n$. 
In azimuthal asymmetries one needs the transverse moments contained in the
correlator
\bea
\label{TransverseMoment}
\Phi_{\partial}^{\alpha\,[\mathcal U]}(x) 
= \int d^2k_\sT\ k_\sT^\alpha\,\Phi^{[\mathcal U]}(x{,}k_\sT)\, .
\eea
The TMD correlator, expanded in distribution functions depending on
$x$ and $k_\sT^2$ contains T-even and T-odd functions, since the
correlator is not T-invariant, which is attributed to the gauge
link that depending on the process, accounts for specific initial 
and/or final state interactions depending on the color flow in the process.
For the collinear case, the link structure becomes unique in the case
of integration over $k_\sT$ (Eq.~\ref{colcorrelator}). For spin 0 and
spin 1/2 the quark and gluon correlators that appear at leading order
in high energy processes contain only T-even operator combinations.
Evaluated between plane waves one only finds T-even functions depending
on $x$ in the parameterization.

For the collinear weighted case, the transverse moments in 
Eq.~\eqref{TransverseMoment} one retains a nontrivial link-dependence
that prohibits the use of T-invariance as a constraint. It is possible,
however, to decompose the weighted quark (and also gluon) correlators as
\begin{equation}\label{DECOMPOSITION}
\Phi_\partial^{\alpha\,[\mathcal U]}(x)
=\tilde \Phi_\partial^{\alpha}(x)
+C_G^{[\mathcal U]}\,\pi\Phi_G^{\alpha}(x,x)\, ,
\end{equation}
with calculable process-dependent gluonic pole factors $C_G^{[\mathcal U]}$
and process (link) independent correlators $\tilde\Phi_\partial$ and
$\Phi_G$. The correlator $\tilde\Phi_\partial$ contains the T-even operator 
combination, while $\Phi_G$ contains the T-odd operator combination.
The latter is precisely the soft limit $x_1\rightarrow 0$ of
a quark-gluon correlator $\Phi_G(x,x_1)$ of the type
\bwt\bea
\label{GP}
\Phi_G^\alpha(x,x{-}x_1)
=n_\mu
\int\frac{d(\xi{\cdot}P)}{2\pi}\frac{d(\eta{\cdot}P)}{2\pi}\ 
e^{ix_1(\eta\cdot P)}e^{i(x-x_1)(\xi\cdot P)}\,
\langle P|\,\overline\psi(0)\,U_{[0;\eta]}^n\,gG^{\mu\alpha}(\eta)\,
U_{[\eta;\xi]}^n\,\psi(\xi)\,|P\rangle\,\big\rfloor_{\text{LC}}\ ,
\eea
The universal T-odd distribution functions in the parameterization 
of $\Phi_G(x,x)$ appear in T-odd observables such as single spin 
asymmetries with the specific gluonic pole factors from 
Eq.~\ref{DECOMPOSITION}.   

The situation for fragmentation functions is different. The TMD fragmentation
correlator depending on the collinear and transverse components of the
quark momentum, $k = \tfrac{1}{z}\,P + k_\sT + \sigma\,n$,
is given by~\cite{Boer:2003cm}
\bea
\Delta^{[\mathcal U]}_{ij}(z,k_\sT)&=&
\sum_X\int\frac{d(\xi\cdot P_h)\,d^2\xi_\sT}{(2\pi)^3}\ e^{i\,k\cdot\xi}
\langle 0 |\mathcal U_{[0,\xi]}\psi_i(\xi)|P,X\rangle
\langle P,X|\bar{\psi}_j(0)|0\rangle |_{LF}\,  .
\label{TMDFF}
\eea
The collinear, $k_\sT$-integrated correlator
\bea
\Delta(z)&=&\int d^2k_\sT
 \Delta^{[\mathcal U]}(z,k_\sT)
=\sum_X\int\frac{d(\xi\cdot P)}{2\pi}\ e^{i\,z^{-1}(\xi\cdot P)}
\langle 0 |\mathcal U^n_{[0,\xi]}\psi_i(\xi)|P,X\rangle
\langle P,X|\bar{\psi}_j(0)|0\rangle |_{LC}\,  ,
\eea
only contains a T-even operator combination. Nevertheless one could
in principle have T-even and T-odd fragmentation functions depending
on $z$ since the hadronic state $\vert P,X\rangle$ is an out-state,
which is not T-invariant. For spin 0 and spin 1/2 hadrons no T-odd
function appear at leading twist because of other constraints. 
At sub-leading twist they do appear~\cite{Jaffe:1993xb}.  

In the transverse moments obtained after $k_\sT$-weighting,
\bea
\Delta^{\alpha\,[\mathcal U]}_{\partial}(z)
&=&\int d^2k_\sT\ k_\sT^\alpha \Delta^{[\mathcal U]}(z,k_\sT)
=\tilde{\Delta}_\partial^\alpha\left(\tfrac{1}{z}\right)
+C_G^{[\mathcal U]}
\,\pi\Delta_G^\alpha\left(\tfrac{1}{z},\tfrac{1}{z}\right).
\label{decompfrag}
\eea
the two link independent correlators $\tilde\Delta_\partial$
and $\Delta_G$ contain again a T-even and T-odd operator combination,
respectively. The gluonic pole correlator is again the soft limit,
$z_1^{-1} = x_1 \rightarrow 0$, of the quark-gluon correlator 
\bea
\Delta_{G\,ij}^\alpha\left(x,x-x_1\right)
=\sum_X \int\frac{d(\xi{\cdot}P)}{2\pi}\frac{d(\eta{\cdot}P)}{2\pi}\,
e^{i\,x_1(\eta\cdot P)}e^{i\,(x-x_1)(\xi\cdot P)}\,
\langle 0 | \mathcal U^n_{[0,\eta]}\, gG^{n\alpha}(\eta)
\,\mathcal U^n_{[\eta,\xi]}\psi_i(\xi)|P,X\rangle
\langle P,X|\overline{\psi}_j(0)|0\rangle\Bigg|_{LC} .
\nn
\label{GLa}
 \eea\ewt
Because of the appearance of hadronic states $\vert P,X\rangle$,
each of correlators in Eq.~\ref{decompfrag} contains in principle 
T-even and T-odd functions.
Rather than having a doubling of T-odd functions, we will
show in a spectator model approach that $\Delta_G(x,x)$ = 0,
which implies that T-odd fragmentation functions in the
transverse moments only come from $\tilde\Delta_\partial$,
which appear with a universal strength (no gluonic pole factors).
We will show this in a spectator approach starting with
the collinear quark-gluon correlators in Eqs.~\ref{GP} and~\ref{GLa}
rather than the model 
approaches~\cite{Bacchetta:2002tk,Ji:2002aa,Gamberg:2003ey,Gamberg:2003eg,Bacchetta:2003xn,Bacchetta:2003rz,Lu:2004hu,Amrath:2005gv,Bacchetta:2007wc,Gamberg:2007gb}
that looked at the  transverse momentum dependent 
quark correlators in Eqs~\ref{TMDDF} and \ref{TMDFF}.

\section{The spectator model approach}

In a typical spectator model approach to distribution or fragmentation
correlators one considers a spectator with mass $M_s$. The 
result for the cut, but untruncated, diagrams, such as in 
Figs.~\ref{qqspec-1} and \ref{qqspec-2} are of the form
\bea
\Phi(x,k_\sT)\hspace{-.2cm} &\sim &\hspace{-.2cm}\int d(k\cdot P)
\frac{F(k^2,k\cdot P)}{(k^2-m^2+i\epsilon)^2}
\delta\left( (k-P)^2 - M_s^2\right),
\nn
\label{basic}
\eea
where  $F(k^2,k\cdot P)$  contains the numerators of propagators and/or
traces of them in the presence of Dirac Gamma matrices, as well as 
the vertex form factors (see for example~\cite{Jakob:1997wg}).
The explicit momenta, using light-cone coordinates $[p^-,p^+,p_\sT]$
as discussed in the beginning of the previous section, are
\bea
&& 
P = \Bigl[\frac{M^2}{2}, 1, 0_\sT\Bigr],
\\&&
P-k = \Bigl[\frac{M_s^2-k_\sT^2}{2(1-x)}, 1-x, -k_\sT\Bigr], 
\\&&
k = \Bigl[\frac{(1-x)M^2-M_s^2+k_\sT^2}{2(1-x)}, x, k_\sT\Bigr] .
\eea
In the above the delta function constraint in Eq.~\ref{basic} has
been implemented. One finds that the numerator
$F(k^2,k\cdot P) = F(x,k_\sT^2)$ and hence
\bea
\Phi(x,k_\sT) \sim 
\ \frac{(1-x)^2\,F(x,k_\sT)}{\left(\mu^2(x)-k_\sT^2\right)^2},
\label{qqspec}
\eea
with
\bea
\mu^2(x) = x\,M_s^2+(1-x)\,m^2-x(1-x)\,M^2.
\label{mu}
\eea
Note that $k_\sT^2 = -\bm k_\sT^2 \le 0$.
The details of the numerator function depend on the details of the 
model, including the vertices, polarization sums, etc. These must
be chosen in such a way as to not produce unphysical effects,
such as a decaying proton if $M \ge m+M_s$, thus $m$ in Eq.~\ref{basic}
must represent some constituent mass in the quark propagator, rather
than the bare mass.
The useful feature of the result in Eq.~\ref{qqspec} is its ability to
produce reasonable valence and even sea quark distributions using the
freedom in the model connected to an intuitive picture.
The results for the fragmentation function in the spectator
model is identical upon the substitution of $x = 1/z$.

Next we turn to the same spectral analysis of the gluonic pole
correlator using the picture given in Figs.~\ref{qqGspec-1} for
distribution functions and the picture given in Figs.~\ref{qqGspec-2}
for fragmentation functions. Again, we only need to investigate
one of the cases. 
We parameterize the gluon momentum as
\bea
k_1 = \Bigl[k_1^-,x_1, k_{1\sT}\Bigr],
\eea
where $k_1^- = k_1\cdot P - \tfrac{1}{2}\,x_1\,M^2$ will be the first
component to be integrated over.
The relevant momenta (implementing the on-shell
condition for $P-k$) are
\bwt
\bea
 k-k_1 &=&\hspace{0.05cm}\Bigl[{-}k_1^- +\frac{(1-x)\,M^2-M_s^2+k_\sT^2}{2(1-x)},
x-x_1, k_\sT-k_{1\sT}\Bigr],
\\ 
P-k+k_1 
&=&\Bigl[k_1^- + \frac{M_s^2-k_\sT^2}{2(1-x)},
1-x+x_1, -k_\sT+k_{1\sT}\Bigr],
 \\ 
P-k_1 
&=& \Bigl[{-}k_1^- + \frac{M^2}{2},1-x_1,{-}k_{1\sT}\Bigr].
\eea
The basic result for the quark-gluon correlators 
$\Phi_G(x,x-x_1,k_\sT,k_\sT-k_{1\sT})$
becomes
\bea
\Phi_G &\sim&
\frac{1}{(k^2-m^2)}\Biggl\{
\int\frac{dk_1^-}{2\pi\,i}
\ \frac{F_1\left(k_1^-,x,x_1,k_\sT,k_{1\sT}\right)}{
(k_1^2-m_1^2+i\epsilon)((k-k_1)^2-m^2+i\epsilon)
((P-k+k_1)^2-M_{s1}^2+i\epsilon)}
\nonumber \\&& \mbox{}\hspace{2cm}
+\int\frac{dk_1^-}{2\pi\,i}
\ \frac{F_2\left(k_1^-,x,x_1,k_\sT,k_{1\sT}\right)}{
(k_1^2-m_1^2+i\epsilon)((k-k_1)^2-m^2+i\epsilon)
((P-k_1)^2-M_{s2}^2+i\epsilon)}\Biggr\}
\nonumber \\
&\sim& \frac{1-x}{(\mu^2-k_\sT^2)}\Biggl\{
\int\frac{dk_1^-}{2\pi\,i}
\ \frac{F_1\left(k_1^-,x,x_1,k_\sT^2,k_{1\sT}^2\right)}{
(x_1\,k_1^--A_1+i\epsilon)((x_1-x)\,k_1^--A_2+i\epsilon)
((1-x+x_1)\,k_1^--B_1+i\epsilon)}
\nonumber \\&& \mbox{}\hspace{2cm}
+\int\frac{dk_1^-}{2\pi\,i}
\ \frac{F_2\left(k_1^-,x,x_1,k_\sT^2,k_{1\sT}^2\right)}{
(x_1\,k_1^--A_1+i\epsilon)((x_1-x)\,k_1^--A_2+i\epsilon)
((x_1-1)\,k_1^--B_2+i\epsilon)}\Biggr\},
\label{resultgp}
\eea
\ewt
where, as before,  $F_i\left(k_1^-,x,x_1,k_\sT^2,k_{1\sT}^2\right)$ contain
numerators and vertex functions~\cite{Ji:2002aa,Gamberg:2003ey}.  
We use the quantities,
\bea
2\,A_1 &=& m_1^2 - k_{1\sT}^2,\\
2\,A_2 &=& m^2 - (x-x_1)\,M^2 
+\frac{x-x_1}{1-x}\,\left(M_s^2-k_\sT^2\right)
\nn  && \hspace{3.2cm} -(k_\sT-k_{1\sT})^2, \\
2\,B_1 &=& M_{s1}^2  
-\frac{1-x+x_1}{1-x}\,\left(M_s^2-k_\sT^2\right)
\nn  &&\hspace{3.2cm} -(k_\sT-k_{1\sT})^2,\\
2\,B_2 &=& M_{s2}^2 - (1-x_1)\,M^2 - k_{1\sT}^2\, .
\eea
These quantities depend on spectator masses, momentum fractions ($x$
and $x_1$) and the transverse momenta ($k_\sT$ and $k_{1\sT}$).
Besides the spectator mass $M_s$, two additional spectator masses $M_{s1}$
and $M_{s2}$ appear (see Figs.~\ref{qqGspec-1}(b) and (c) or 
Figs.~\ref{qqGspec-2}(b) and (c) and compare with the starting 
expression in Eq.~\ref{resultgp}). The quantity $\mu^2$ is the same
one as given in Eq.~\ref{mu}.
Assuming that the numerator does not grow with $k_1^-$ one can
easily perform the $k_1^-$ integrations. We will for simplicity
assume that the $F_i$ are independent of $k_1^-$, in which case we
obtain for $x \ge x_1 \ge 0$ (see Appendix),
\bwt
\bea
\Phi_G(x,x-x_1) &=& \int d^2k_\sT\,d^2k_{1\sT}
\nonumber \\&& \mbox{}\hspace{-0.5cm}
\Biggl\{ \frac{(1-x)\,F_1(x,x_1,k_\sT,k_{1\sT})}{\bigl(\mu^2-k_\sT^2\bigr)
\bigl(x\,A_1+x_1\,(A_2-A_1)\bigr)
\bigl(x_1\,(B_1-A_2)-x\,B_1-(1-x)\,A_2\bigr)
\bigl((1-x)\,A_1+x_1\,(A_1-B_1)\bigr)}
\nonumber \\&& \mbox{}\hspace{3.0cm}
{}\times\Biggl[\bigl(x_1\,(B_1-A_2)-x\,B_1-(1-x)\,A_2\bigr)\,x_1\theta(x_1)
\nonumber \\&& \mbox{}\hspace{3.5cm}
{}+\bigl((1-x)\,A_1+x_1\,(A_1-B_1)\bigr)\,(x_1-x)\theta(x_1-x)
\nonumber \\&& \mbox{}\hspace{3.5cm}
{}+\bigl(x\,A_1+x_1\,(A_2-A_1)\bigr)\,(1-x+x_1)\theta(1-x+x_1)\Biggr]
\nonumber \\&&\mbox{}\hspace{-0.5cm}
{}+\frac{(1-x)\,F_2(x,x_1,k_\sT,k_{1\sT})}{\bigl(\mu^2-k_\sT^2\bigr)
\bigl(x\,A_1+x_1\,(A_2-A_1)\bigr)
\bigl(A_2-x\,B_2+x_1\,(B_2-A_2)\bigr)
\bigl(x_1\,(A_1-B_2)-A_1\bigr)}
\nonumber \\&& \mbox{}\hspace{3.0cm}
{}\times\Biggl[\bigl(A_2-x\,B_2+x_1\,(B_2-A_2)\bigr)\,x_1\theta(x_1)
\nonumber \\&& \mbox{}\hspace{3.5cm}
{}+\bigl(x_1\,(A_1-B_2)-A_1\bigr)\,(x_1-x)\theta(x_1-x)
\nonumber \\&& \mbox{}\hspace{3.5cm}
{}+\bigl(x\,A_1+x_1\,(A_2-A_1)\bigr)\,(x_1-1)\theta(x_1-1)\Biggr]
\Biggr\}
\eea
Taking the limit $x_1 \rightarrow 0$ we get the gluonic pole correlators,
for distribution functions ($0\le x \le 1$),
\bea
\Phi_G(x,x) 
&=& - \int d^2k_\sT\,d^2k_{1\sT}
\ \frac{(1-x)\,F_1(x,0,k_\sT,k_{1\sT})\theta(1-x)}{
\bigl(\mu^2-k_\sT^2\bigr) \bigl(x\,B_1+(1-x)\,A_2\bigr)\,A_1}\, ,
\eea
\ewt
and for fragmentation functions ($x = 1/z \ge 1$)
\bea
\Delta_G(x,x) &=& 0\, .
\eea
We note that this result depends on the assumption that the numerator
does not grow with $k_1^-$. If this is the case one would find integrals
of the type $\Theta^1_{111}$ (also given in the Appendix) 
rather than those of the type $\Theta^0_{111}$
and one does not get the required $x_1\,\theta(x_1)$ behavior in the
calculation. In  models, terms proportional to $k_1^- \sim k_1\cdot P$ 
may easily arise from numerators of fermionic 
propagators~\cite{Gamberg:2006ru,Gamberg:2007gb}
which in turn may easily be suppressed by  form factors at the
vertices. To prove a proper behavior within QCD one would need to study the
fully unintegrated correlators such as e.g.\ in 
Ref.~\cite{Collins:2007ph}
and show that they fall off sufficiently fast as a function of $k_1\cdot P$.

\section{Conclusions\label{Conclusion}}

In this work, we have investigated the gluonic pole contributions to the
distribution and fragmentation functions. Instead of doing a
quantitative analysis involving details of a phenomenological model, we limit
ourselves to a spectral analysis within the spectator framework,
in order to understand the basic
features of these quantities. The advantage is that we are able to investigate
only the soft parts at tree level and take the zero momentum limit of the gluon
involved. We simply assume that masses and vertices do not spoil our
analysis, which implies limits on the mass distributions of the
spectators, use of vertices that cancel the bare-mass poles in the quark 
and gluon propagators and behavior of vertices that assures sufficient
convergence of integrations.
We find that under realistic assumptions, the gluonic pole
contributions for  fragmentation correlators vanish whereas 
these contributions do not vanish for distribution correlators. 
The result for fragmentation correlators at nonzero gluon momentum 
is nonzero. We stress that this certainly is not yet the full proof that 
gluonic pole  matrix elements vanish in the case of fragmentation. 
However, we consider this analysis as a step towards such a  proof and 
the possible direction to
obtain such a proof by only considering the appropriate color gauge-invariant
soft matrix elements.
Such a proof is important as it eliminates a whole class of matrix
elements parameterized in terms of T-odd fragmentation functions besides
the T-odd fragmentation functions 
in the parameterization of the two-parton correlators. 
For instance, only one of the contributions to the spin 
asymmetries considered for jet-hyperon production in 
Ref.~\cite{Boer:2007nh} remains.  Moreover,the 
remaining fragmentation functions appear with the 
standard partonic cross section, so no gluonic pole cross sections need to be
considered for fragmenting final-state partons, limiting these considerations
to the distribution functions involving initial-state partons.

As mentioned, the results in this paper may point the way to find
a full proof of $\Delta_G(x,x) = 0$. The approach taken here to look
at tree-level three-parton correlators, also can be used for explicit
model calculations for T-odd distribution functions originating from
gluonic pole matrix elements and the investigation of 
their effects in single spin asymmetries. 

\begin{acknowledgments}
\vskip -.25cm
This work was initiated at the WHEPP-X workshop in Chennai (January 2008).
We thank Daniel Boer for  useful discussions. 
L.G. acknowledges support from  U.S. Department of Energy under contract
DE-FG02-07ER41460. 
\end{acknowledgments}
\vspace{-1cm}
\bwt
\appendix
\section{Useful Integrals\label{Theta-functions}}
Often it is useful to attack integrals containing propagator poles
via light-cone variables, leading to integrals of the type
\bea
\Theta_{n_1n_2\ldots}^{m}(x_1,x_2,\ldots) =
\int\frac{d\alpha}{2\pi\,i}
\ \frac{\alpha^m}{(\alpha x_1-1+i\epsilon)^{n_1}
(\alpha x_2-1+i\epsilon)^{n_2}\ldots},
\eea
for which easy reduction rules exist~\cite{Belitsky:1997ay}.  
We need specifically
\bea
&&
\Theta_{11}^0(x_1,x_2) =
\frac{\theta(x_1)\theta(-x_2) 
-\theta(-x_1)\theta(x_2)}{x_1-x_2} 
= \frac{\theta(x_1)-\theta(x_2)}{x_1-x_2} ,
\\ &&
\Theta_{111}^0(x_1,x_2,x_3) =
\frac{x_2}{(x_1-x_2)}\,\Theta_{11}^0(x_2,x_3)
-\frac{x_1}{(x_1-x_2)}\,\Theta_{11}^0(x_1,x_3)
\\ && \mbox{}\hspace{2cm} =
\frac{(x_2-x_3)\,x_1\,\theta(x_1)+(x_3-x_1)\,x_2\,\theta(x_2)
+(x_1-x_2)\,x_3\,\theta(x_3)}{(x_1-x_2)(x_2-x_3)(x_3-x_1)},
\\ &&
\Theta_{111}^1(x_1,x_2,x_3) =
\frac{1}{(x_1-x_2)}\left[\Theta_{11}^0(x_2,x_3)
-\Theta_{11}^0(x_1,x_3)\right]
\\ && \mbox{}\hspace{2cm} =
\frac{(x_2-x_3)\,\theta(x_1)+(x_3-x_1)\,\theta(x_2)
+(x_1-x_2)\,\theta(x_3)}{(x_1-x_2)(x_2-x_3)(x_3-x_1)}.
\eea
Including arbitrary pole positions, one finds
\bea
&&
\int\frac{d\alpha}{2\pi\,i}
\ \frac{1}{(\alpha x_1-A_1+i\epsilon)(\alpha x_2-A_2+i\epsilon)}
\nonumber\\ &&\mbox{} \hspace{1cm} \mbox{}
= \frac{x_1-x_2}{x_1A_2-x_2A_1}
\,\Theta_{11}^0(x_1,x_2)
\\ &&\mbox{} \hspace{1cm} \mbox{}
= \frac{\theta(x_1)-\theta(x_2)}{x_1A_2-x_2A_1}
= \frac{\theta(x_1)\,\theta(-x_2)-\theta(-x_1)\,\theta(x_2)}{x_1A_2-x_2A_1},
\\ &&
\int\frac{d\alpha}{2\pi\,i}
\ \frac{1}{(\alpha x_1-A_1+i\epsilon)(\alpha x_2-A_2+i\epsilon)
(\alpha x_1-A_3+i\epsilon)}
\nonumber\\ &&\mbox{} \hspace{1cm} \mbox{}
= \frac{1}{x_1A_2-x_2A_1}\left[
x_2\,\frac{x_2-x_3}{x_2A_3-x_3A_2}
\,\Theta_{11}^0(x_2,x_3)
-x_1\,\frac{x_1-x_3}{x_1A_3-x_3A_1}
\,\Theta_{11}^0(x_1,x_3)\right]
\\ &&\mbox{}\hspace{1cm} \mbox{}
= \frac{(x_2A_3-x_3A_2)\,x_1\,\theta(x_1)
+ (x_3A_1-x_1A_3)\,x_2\,\theta(x_2)
+ (x_1A_2-x_2A_1)\,x_3\,\theta(x_3)}
{(x_1A_2-x_2A_1) (x_2A_3-x_3A_2) (x_3A_1-x_1A_3)},
\\ &&
\int\frac{d\alpha}{2\pi\,i}
\ \frac{\alpha}{(\alpha x_1-A_1+i\epsilon)(\alpha x_2-A_2+i\epsilon)
(\alpha x_1-A_3+i\epsilon)}
\nonumber\\ &&\mbox{} \hspace{1cm} \mbox{}
= \frac{1}{x_1A_2-x_2A_1}\left[
A_2\,\frac{x_2-x_3}{x_2A_3-x_3A_2}
\,\Theta_{11}^0(x_2,x_3)
-A_1\,\frac{x_1-x_3}{x_1A_3-x_3A_1}
\,\Theta_{11}^0(x_1,x_3)\right]
\\ &&\mbox{}\hspace{1cm} \mbox{}
= \frac{(x_2A_3-x_3A_2)\,\theta(x_1)
+ (x_3A_1-x_1A_3)\,\theta(x_2)
+ (x_1A_2-x_2A_1)\,\theta(x_3)}
{(x_1A_2-x_2A_1) (x_2A_3-x_3A_2) (x_3A_1-x_1A_3)}.
\eea
\ewt

\bibliographystyle{apsrev}

\bibliography{references}

\end{document}